\documentclass[aps,pre,twocolumn,superscriptaddress,fleqn,tbtags,onecolumn%
]{revtex4}%

\usepackage[english]{babel}%
\usepackage{amsmath,amssymb}%
\usepackage{graphicx}%
\usepackage{grffile}%
\usepackage{ifthen}%
\usepackage[squaren]{SIunits}%
\usepackage{color}%
\usepackage{hyperref}%
\usepackage{dcolumn}
\usepackage{bm}
\usepackage[section]{placeins}
\usepackage{xr}
\newcommand{\Bstate}{B}
\newcommand{\Sstate}{S}
\newcommand{\Mstate}{M}

\newcommand{\Q}{{\mathcal{Q}}}%
\newcommand{\QM}{\Q_\mathrm{\Mstate}}%

\newcommand{\Z}{{\mathcal{Z}}}%
\newcommand{\ZB}{\Z_\mathrm{\Bstate}}%
\newcommand{\ZS}{\Z_\mathrm{\Sstate}}%
\newcommand{\ZM}{\Z_\mathrm{\Mstate}}%

\newcommand{\V}[2]{V_{#1#2}}%
\newcommand{\VBS}{\V{\mathrm{\Bstate}}{\mathrm{\Sstate}}}%
\newcommand{\VBM}{\V{\mathrm{\Bstate}}{\mathrm{\Mstate}}}%
\newcommand{\VSB}{\V{\mathrm{\Sstate}}{\mathrm{\Bstate}}}%
\newcommand{\VSM}{\V{\mathrm{\Sstate}}{\mathrm{\Mstate}}}%
\newcommand{\VMB}{\V{\mathrm{\Mstate}}{\mathrm{\Bstate}}}%
\newcommand{\VMS}{\V{\mathrm{\Mstate}}{\mathrm{\Sstate}}}%
\newcommand{\Vii}{V_{ii}}%
\newcommand{\Vij}{V_{ij}}%

\newcommand{\FEi}{E_{i}}%

\newcommand{\g}{g}%
\newcommand{\gB}{\g_{\mathrm{\Bstate}}}%
\newcommand{\gBo}{\g_{\mathrm{\Bstate}}^{0}}%
\newcommand{\gS}{\g_{\mathrm{\Sstate}}}%
\newcommand{\gSo}{\g_{\mathrm{\Sstate}}^{0}}%
\newcommand{\gM}{\g_{\mathrm{\Mstate}}}%
\newcommand{\gMo}{\g_{\mathrm{\Mstate}}^{0}}%
\newcommand{\gi}{\g_{i}}%
\newcommand{\gio}{\g_{i}^{0}}%
\newcommand{\giWLC}{\g_{i}^{\mathrm{WLC}}}%
\newcommand{\gistretch}{\g_{i}^{\mathrm{stretch}}}%

\newcommand{\Fc}{\F_{\mathrm{c}}}
\newcommand{\F}{F}
\newcommand{\FWLC}{\F^{\mathrm{WLC}}}
\newcommand{\GFEGC}{\Phi}
\newcommand{\GFE}{\mathcal G}
\newcommand{\Li}[2]{\mathrm{Li}_{#1}\left(#2\right)}%
\newcommand{\Nbb}{\mathbb{N}}
\newcommand{\Tc}{T_{\mathrm{c}}}
\newcommand{\cf}{\emph{cf.\ }}%
\newcommand{\ci}{\mathrm{i}}
\newcommand{\dd}{\mathrm d}%
\newcommand{\dotproduct}{\cdot}                           
\newcommand{\eqcomment}[1]{\quad\quad\text{#1}}%
\newcommand{\eqspace}{\;}%
\newcommand{\eqs}{eqs.}%
\newcommand{\eq}{eq.}%
\newcommand{\e}{\mathrm{e}}%
\newcommand{\fig}{fig.}%
\newcommand{\ie}{\emph{i.\,e.\ }}%
\newcommand{\kB}{\mathrm{k_B}}%
\newcommand{\kronecker}[1]{\delta_{#1}}%
\newcommand{\lambdab}{\lambda_{\mathrm{b}}}
\newcommand{\lambdad}{\lambda_{\mathrm{d}}}
\newcommand{\lambdap}{\lambda_{\mathrm{p}}}
\newcommand{\loopexponent}{c}%
\newcommand{\mat}[1]{\bm{\mathrm{#1}}}
\newcommand{\openendexponent}{{c}}%

\newcommand{\persistencelengthB}{\persistencelength_{\mathrm{\Bstate}}}%
\newcommand{\persistencelengthM}{\persistencelength_{\mathrm{\Mstate}}}%
\newcommand{\persistencelengthS}{\persistencelength_{\mathrm{\Sstate}}}%
\newcommand{\persistencelength}{\xi}%
\newcommand{\scalegSo}{\tau_{\mathrm{\Sstate}}}%
\newcommand{\segmentlengthB}{\segmentlength_{\mathrm{\Bstate}}}%
\newcommand{\segmentlengthM}{\segmentlength_{\mathrm{\Mstate}}}%
\newcommand{\segmentlengthS}{\segmentlength_{\mathrm{\Sstate}}}%
\newcommand{\segmentlength}{l}%
\newcommand{\stretchmodulusB}{\stretchmodulus_{\mathrm{\Bstate}}}%
\newcommand{\stretchmodulusM}{\stretchmodulus_{\mathrm{\Mstate}}}%
\newcommand{\stretchmodulusS}{\stretchmodulus_{\mathrm{\Sstate}}}%
\newcommand{\stretchmodulus}{\kappa}%
\newcommand{\transpose}[1]{#1^{\mathrm{T}}}
\newcommand{\via}{\emph{via} }%
\renewcommand{\vec}[1]{\bm{\mathrm{#1}}}
\addunit{\molar}{M}%
\addunit{\calory}{cal}%
\usepackage{soul}
\newcommand{\DELETEcdot}{\cdot}
\newcommand{\trecdot}{\cdot}
\newcommand{\ResubNew}[1]{#1}

\begin{document}
\title{\ResubNew{A three-state model with loop entropy for the over-stretching transition
    of DNA}}

\author{Thomas R. Einert} \email[Corresponding author:]{ Physik Department, Technische
  Universit\"at M\"unchen, James-Franck-Stra\ss e, 85748 Garching, Germany, Tel.:
  +49-89-28914337, Fax: +49-89-28914642, E-mail: einert@ph.tum.de} \affiliation{Physik
  Department, Technische Universit\"at M\"unchen, 85748 Garching, Germany}%
\author{Douglas B. Staple} \affiliation{Department of Physics and Atmospheric Science,
  Dalhousie University, Halifax, B3H 3J5 Canada}\affiliation{Max-Planck-Institut f\"ur
  Physik komplexer Systeme, 01187 Dresden, Germany}%
\author{Hans-J\"urgen Kreuzer} \affiliation{Department of Physics and Atmospheric Science,
  Dalhousie University, Halifax, B3H 3J5 Canada}%
\author{Roland R. Netz} \email[E-mail: ]{netz@ph.tum.de} \affiliation{Physik Department,
  Technische Universit\"at M\"unchen, 85748 Garching, Germany}

\date{\today}
\begin{abstract}
  We introduce a three-state model for a single DNA chain under tension that distinguishes
  between \Bstate-DNA, \Sstate-DNA, and \Mstate{} (molten or denatured) segments and at the
  same time correctly accounts for the entropy of molten loops, characterized by the
  exponent~$\loopexponent{}$ in the asymptotic expression $S\sim -\loopexponent{} \ln n$
  for the entropy of a loop of length $n$. Force extension curves are derived exactly
  employing a generalized Poland-Scheraga approach and compared to experimental
  data. Simultaneous fitting to force-extension data at room temperature and to the
  denaturation phase transition at zero force is possible and allows to establish a global
  phase diagram in the force-temperature plane. Under a stretching force, the effects of
  the stacking energy, entering as a domain-wall energy between paired and unpaired bases,
  and the loop entropy are separated. Therefore we can estimate the loop exponent~$\loopexponent{}$ independently from the precise value of the stacking energy. The
  fitted value for $\loopexponent{}$ is small, suggesting that nicks dominate the
  experimental force extension traces of natural DNA.
\end{abstract}

\maketitle
\section{Introduction}
\label{sec:introduction}

DNA continuously stays in focus of polymer scientists due to its unique mechanical and
structural properties.  In particular the possibility to trigger phase transformations in
this one-dimensional system has intrigued theorists from different
areas~\cite{Poland1966}. \ResubNew{In fact, the thermal denaturation or melting transition
  of DNA was shown to correspond to a true phase transition, brought about by a
  logarithmic contribution to the configurational entropy of molten loops or bubbles,
  $S\sim -\loopexponent{} \ln n$, as a function of the loop size~$n$~\cite{Poland1966a}.}
The value of the exponent~$\loopexponent{}$ is crucial since it determines the resulting
transition characteristics.  For $\loopexponent{}=3/2$, the value for a phantom chain
without self-avoidance, the transition is continuous, while self-avoidance increases $\loopexponent{}$ slightly beyond the threshold $\loopexponent{}=2$ above which the
transition becomes discontinuous~\cite{Kafri2002,Poland1966a}.  A distinct mechanism for
transforming DNA involves the application of an extensional force.  For forces around
$\F\approx\unit{65}{\pico\newton}$ DNA displays a highly cooperative transition and its
contour length increases by a factor of roughly 1.7 to 2.1 over a narrow force
range~\cite{Smith1996,Bensimon1995,Cluzel1996}.  \ResubNew{These experiments sparked a
  still ongoing debate on whether this over-stretching transition produces a distinct DNA
  state, named \Sstate-DNA, or merely the denatured state under external tension.
  According to the first view \Sstate-DNA is a highly stretched state with paired bases
  but disrupted base stacking
  \cite{Cocco2004,Leger1999,Whitelam2008,Konrad1996,Kosikov1999,Li2009,Lebrun1996}.  In
  the other view the over-stretched state consists of two non-interacting strands
  \cite{Williams2001a,Rouzina2001,Shokri2008,Mameren2009}.  Evidence for the existence of
  a distinct \Sstate-state comes from theoretical models~\cite{Cocco2004,Whitelam2008},
  molecular dynamics simulations~\cite{Konrad1996,Kosikov1999,Lebrun1996,Li2009}} and from
AFM experiments of
\citeauthor{Rief1999}~\cite{Rief1999,ClausenSchaumann2000,Krautbauer2000} where in
addition to the over-stretching transition a second weak transition at forces between
\unit{150}{\pico\newton} and \unit{300}{\pico\newton} is discerned, which has been
interpreted as a force induced melting of the \Sstate-state.  The critical force of both
transitions depends on the actual sequence~\cite{ClausenSchaumann2000} and the salt
concentration~\cite{Baumann1997}, but the interpretation of the second transition is
complicated by the occurrence of pronounced hysteresis effects that depend on various
parameters such as pulling velocity, salt concentration, or presence of co-solutes such as
cisplatin~\cite{ClausenSchaumann2000,Krautbauer2000}. \ResubNew{On the other hand, support
  for the view according to which \Sstate-DNA is not a distinct state comes from
  theoretical models~\cite{Rouzina2001,Williams2001a}, simulations~\cite{Piana2005} and
  recent experiments by \citet{Shokri2008} and \citet{Mameren2009}.}

\ResubNew{Apart from simulations \cite{Konrad1996,Kosikov1999,Lebrun1996,Li2009}, existing
  theoretical works that grapple with experimental force traces or DNA melting fall into
  three categories with increasing computational complexity, for reviews see
  refs.~\cite{Cocco2002a,Peyrard2008,Wartell1985}. In the first group are Ising-like
  models for DNA under tension which give excellent fitting of the over-stretching
  transition but by construction cannot yield the denaturing transition
  \cite{Cizeau1997,Ahsan1998,Storm2003a,Rouzina2001}. The work of \citet{Marko1998} is
  similar but employs a continuous axial strain variable. In the second group are models
  that include a logarithmic entropy contribution of molten loops in the spirit of the
  classical model by \citet{Poland1966a}
  \cite{Hanke2008,Rudnick2008,Whitelam2008,Garel2004,Kafri2002,Blake1999,Carlon2002}.
  This gives rise to effectively long-ranged interactions between base pairs and thus to a
  true phase transition. The third group consists of models which explicitly consider two
  strands \cite{Rahi2008a,Palmeri2007,Jeon2006,Peyrard2004,Cule1997}. Those models thereby
  account for the configurational entropy of loops -- at the cost of considerable
  calculational efforts -- and correspond to loop exponents $\loopexponent{}=d/2$ in the
  absence of self-avoidance effects, where the dimensionality of the model is $d = 3$ for
  ref.~\cite{Rahi2008a,Palmeri2007,Jeon2006} and $1$ for ref.~\cite{Peyrard2004,Cule1997}.
  All these above mentioned works consider only two different base states (paired versus
  unpaired) and thus do not allow to distinguish between \Bstate-DNA, \Sstate-DNA and
  denatured bases.  Recently, three-state models were introduced that yield very good fits
  of experimental force traces at ambient temperatures.  However, in previous analytic
  treatments of such three-state models~\cite{Cocco2004,Ho2009}, the loop entropy was
  neglected and therefore the temperature-induced denaturation in the absence of force
  cannot be properly obtained, while the loop entropy was included in a simulation study
  where most attention was given to dynamic effects~\cite{Whitelam2008}.}

\ResubNew{In this paper we combine the Poland-Scheraga formalism with a three-state
  transfer matrix approach which enables us to include three distinct local base pairing
  states and at the same time to correctly account for long-ranged interactions due to the
  configurational entropy of molten DNA bubbles.  Our approach thus allows for a
  consistent description of thermal denaturation and the force induced
  \Bstate\Sstate-transition within one framework and yields the global phase diagram in
  the force-temperature plane.  We derive a closed form expression for the partition
  function of three-state DNA under tension. This allows to systematically investigate the
  full parameter range characterizing the three states and the DNA response to temperature
  and external force.  In our model we allow for the existence of \Sstate-DNA but stress
  that the actual occurrence of \Sstate-DNA is governed by the model parameters.  By
  assuming such a general point of view, our work is able to shed new light on the
  question of the existence of \Sstate-DNA.  The extensible worm-like chain model is
  employed for the stretching response of each state.  The loop exponent is found to have
  quite drastic effects on the force extension curve.}  For realistic parameters for the
stacking energy, the experimental force extension curves are fitted best for small loop
exponents~$0\leq\openendexponent{}\leq1$, hinting that the DNA in the experiments
contained nicks.  Loop exponents~$\loopexponent{}>1$, which give rise to a genuine phase
transition, are not compatible with experimental force-distance curves.  Under external
force, the effects of stacking energy and loop exponent are largely decoupled, since the
stacking energy only determines the cooperativity of the \Bstate\Sstate-transition while
the loop exponent influences the second \Sstate\Mstate-transition found at higher forces.
This allows to disentangle these two parameters, in contrast to the denaturation
transition at zero force where the effects of these two parameters are essentially
convoluted. \ResubNew{The precise value of $\loopexponent{}$ is important also from a
  practical point of view, as it impacts the kinetics of DNA
  melting~\cite{Whitelam2008,Bar2007}, which is omnipresent in biological and
  bio-technological processes.}

\section{Three-state model}
\label{sec:three-state-model}

Double-stranded DNA is modeled as a one-dimensional chain with bases or segments that can
be in three different states, namely paired and in the native \Bstate-state, in the paired
stretched \Sstate-state, or in the molten \Mstate-state.  The free energy of a region of
$n$ segments in the same state reads
\begin{equation}
  \label{eq:1}
  \FEi (n,\F) = n\DELETEcdot g_i (\F)  -\kronecker{i,\mathrm{\Mstate}}\DELETEcdot\kB T \ln
  n^{-\loopexponent} \eqspace,
\end{equation}
with $i=\mathrm{\Bstate,\Sstate,\Mstate}$.  The force~$\F$ dependent contribution
\begin{equation}
  \label{eq:2}
  \gi(\F)=\gio + \gistretch(\F) +\giWLC(\F)
\end{equation}
is split into three parts. $\gio$ is a constant that accounts for the base pairing as well
as the difference of reference states of the worm-like stretching energy, \cf supporting
material \eq~\eqref{appendix-eq:30}. \ResubNew{The stacking energy of neighboring bases in
  the same state is absorbed into $\gio$, too, so that the stacking energy will appear
  explicitly only as an interfacial energy~$\Vij$ between two regions which are in a
  different state.}  The second term $\gistretch = - \F^2
\segmentlength_i/(2\DELETEcdot\stretchmodulus_i)$ takes into account stretching along the
contour with $\segmentlength_i$ and $\stretchmodulus_i$ the segmental contour length and
the elastic stretch modulus.  Finally, $\giWLC(\F)$ is the free energy of a worm-like
chain (WLC) in the Gibbs ensemble (constant force $\F$), based on the heuristic relation
between force~$\F$ and projected extension~$x$~\cite{Marko1995}
%
$ \FWLC_i(x) \trecdot \persistencelength_i / \kB T = \bigl(1-
x/(n\segmentlength_i)\bigr)^{-2}/4 + x/(n\segmentlength_i) - 1/4$
%
where $\persistencelength_i$ is the persistence length and $n$ the number of segments. The
Gibbs free energy~$n\DELETEcdot\giWLC(\F)$ of a stretch of $n$ segments is extensive
in~$n$ and follows \via integration, see supporting material
section~\ref{appendix-sec:gibbs-free-energy}.  We note that this is only valid if the
persistence length is smaller than the contour length of a region,
$\persistencelength_i<n\segmentlength_i$, which is a plausible assumption because of the
high domain wall energies.  Likewise, the decoupling of the free energy into contour
stretching elasticity and worm-like chain elasticity is only
approximate~\cite{Netz2001a,Livadaru2003} but quite accurate for our parameter
values~\cite{Odijk1995}: For small force WLC bending fluctuations dominate and the contour
extensibility is negligible, while contour stretching sets in only when the WLC is almost
completely straightened out.  \ResubNew{The last term in \eq~\eqref{eq:1} is the
  logarithmic configurational entropy of a molten loop ($i=\mathrm{\Mstate}$),
  characterized by an exponent~$\loopexponent{}$ \cite{Kafri2002,Hanke2008,Einert2008},
  see supporting material section~\ref{appendix-sec:orig-logar-loop}.} The exponent is
$\loopexponent{}=3/2$ for an ideal polymer~\cite{Gennes1979} and $2.1$ for a self avoiding
loop with two attached helices~\cite{Duplantier1986,Kafri2002}. If the DNA loop contains a
nick the exponent is reduced to $\openendexponent{}=0$ for an ideal polymer and $0.092$
for a self avoiding polymer~\cite{Kafri2002}.  We consider the simple case
$\loopexponent{}=0$, where transfer matrix methods can be used to yield results in the
canonical ensemble with a fixed number of segments~$N$~\cite{Cocco2004}, as well as the
case of finite~$\loopexponent{}$ where we introduce a modified Poland-Scheraga method to
obtain results in the grand canonical ensemble.

\section{Partition function}
\label{sec:poland-scher-appr}
\subsection{Modified Poland-Scheraga approach for $\loopexponent{}\neq 0$ }
\label{sec:gener-form-canon}

\label{sec:grand-canon-part}

The molecule is viewed as an alternating sequence of different regions each characterized
by grand canonical partition functions.  Various techniques for going back to the
canonical ensemble are discussed below.  The canonical partition function of a stretch of
$n$ segments all in state~$i=\text{\Bstate, \Sstate, or \Mstate}$ is
\begin{equation}
  \Q_i(n)=\exp({-\beta\FEi(n)})\label{eq:3}\eqspace,
\end{equation}
where $\beta = (\kB T)^{-1}$ is the inverse thermal energy.  The grand canonical partition
functions are defined as $\Z_i = \sum_{n=1}^\infty\lambda^n \Q_i(n)$ with $\lambda =
\exp(\beta\mu)$ the fugacity and $\mu$ the chemical potential.  The grand canonical
partition function of the whole DNA chain which contains an arbitrary number of
consecutive \Bstate{}, \Sstate{}, and \Mstate{} stretches reads
\begin{equation}
  \label{eq:4}
  \Z = \sum_{k=0}^\infty\transpose{\vec v}\dotproduct(\mat M_{\mathrm{PS}} \mat V_{\mathrm{PS}})^k\mat M_{\mathrm{PS}}\dotproduct\vec v
  =\transpose{\vec v}\dotproduct(\mat{1} - \mat M_{\mathrm{PS}}\mat V_{\mathrm{PS}})^{-1}\mat M_{\mathrm{PS}}\dotproduct\vec v,
\end{equation}
\ResubNew{with the matrices in this Poland-Scheraga approach given by}
\begin{equation}
  \label{eq:5}
  \mat M_{\mathrm{PS}} =\begin{pmatrix}\ZB&0&0\\0&\ZS&0\\0&0&\ZM\end{pmatrix}\eqspace,
  \quad
  \mat V_{\mathrm{PS}}=\begin{pmatrix}
    0&    \e^{-\beta\VBS}&    \e^{-\beta\VBM}\\
    \e^{-\beta\VSB}& 0 &    \e^{-\beta\VSM}\\
    \e^{-\beta\VMB}&    \e^{-\beta\VMS}&    0
  \end{pmatrix}\eqspace,
  \quad
  \vec v =\begin{pmatrix}1\\1\\1\end{pmatrix}
\end{equation}
and where $\mat 1$ is the unity matrix. The energies $V_{ij}$ are the interfacial energies
to have neighboring segments in different states and are dominated by unfavorable base
pair un-stacking.  The diagonal elements of $\mat V_{\mathrm{PS}}$ are zero which ensures
that two neighboring regions are not of the same type and thus prevents double counting.
The explicit form of $\Z$ is given in the supporting material, see
\eq~\eqref{appendix-eq:36}. \ResubNew{The partition function in \eq~\eqref{eq:4} is
  general and useful for testing arbitrary models for the three DNA states as given by the
  different $\Z_i$. This approach is also easily generalized to higher numbers of
  different states.  Using the parameterization \eq~\eqref{eq:1} for vanishing loop
  exponent~$\loopexponent{}=0$ the partition functions of the different regions are given
  by}
\begin{equation}
  \label{eq:6}
  \Z_i=\sum_{n=1}^\infty\lambda^n\Q_i(n)=\frac{\lambda \e^{-\beta \gi }}{1-\lambda\e^{-\beta \gi }}
  \eqspace, \eqcomment{for $\lambda  \e^{-\beta \gi }<1$,}
\end{equation}
$i=\mathrm{\Bstate,\Sstate,\Mstate}$.  Insertion into \eq~\eqref{eq:4} yields
\begin{equation}
  \label{eq:7}
  {\Z}_{\loopexponent=0} = \frac{a_1 \lambda + a_2 \lambda^2 + a_3\lambda^3}{a_4 + a_5 \lambda + a_6 \lambda^2 + a_7\lambda^3}\eqspace.
\end{equation}
which is a rational function of the fugacity $\lambda$, whose coefficients $a_i$
--~determined by \eqs~\eqref{eq:4} and~\eqref{eq:6}~-- are smooth functions of the
force~$\F$ and the temperature~$T$.  For $\loopexponent{}\neq0$ the partition function of
a molten stretch is modified to
\begin{equation}
  \label{eq:8}
  \begin{split}
    \ZM& = \sum_{n=1}^\infty\lambda^n\QM(n) =
    \sum_{n=1}^\infty \lambda^n\left(\e^{-\beta \gM }\right)^n \frac1{n^\loopexponent}\\
    & = \Li{\loopexponent}{\lambda\e^{-\beta \gM }} \eqspace, \eqcomment{for $\lambda
      \e^{-\beta \gM }<1$}
  \end{split}
\end{equation}
where $\Li{\loopexponent}{z} = \sum_{n=1}^\infty z^nn^{-\loopexponent}$ for $z<1$ is the
polylogarithm~\cite{Erdelyi1953} and exhibits a branch point at $z=1$.  The functional
form of the grand canonical partition function for $\loopexponent{}\neq0$ reads
\begin{equation}
  \label{eq:9}
  {\Z}_{\loopexponent\neq0} = \frac{
    b_0 \lambda + b_1\lambda^2 + b_2 \Li{\loopexponent}{\lambda /\lambdab} +
    b_3 \lambda \Li{\loopexponent}{\lambda /\lambdab} +
    b_4\lambda^2 \Li{\loopexponent}{\lambda /\lambdab}
  }{
    b_5 + b_6 \lambda + b_7\lambda^2 + b_8 \lambda \Li{\loopexponent}{\lambda /\lambdab} +
    b_9\lambda^2 \Li{\loopexponent}{\lambda /\lambdab}
  }
  \eqspace,
\end{equation}
where $\lambdab=\e^{\beta \gM }$ denotes the position of the branch point and the
coefficients $b_i$, determined by \eqs~\eqref{eq:4}, \eqref{eq:6} and~\eqref{eq:8}, are
smooth functions of $\F$ and~$T$.

\label{sec:canon-part-funct}

\ResubNew{The grand canonical ensemble where $N$, the total number of segments fluctuates,
  does not properly describe a DNA chain of fixed length.  We therefore have to
  investigate the back-transformation into the canonical ensemble where the number of
  segments $N$ is fixed. For the back-transformation there are three options:}
\subsubsection{Calculus of residues route:}
\label{sec:calc-resid-route}
The grand-canonical partition function $\Z(\lambda) = \sum_{N=1}^\infty \lambda^N \Q(N)$
can be viewed as a Laurent series, the coefficients of which are the canonical partition
functions $\Q(N)$ determined exactly by
\begin{equation}
  \label{eq:10}
  \Q(N) = \frac1{2\pi\ci}\oint_{\mathcal C}\frac{\Z(\lambda)}{\lambda^{N+1}}\dd\lambda\eqspace.
\end{equation}
The contour $\mathcal C = \lambda_0 \e^{2 \pi\ci t}$, $0\leq t\leq1$, is a circle in the
complex plane around the origin with all singularities of $\Z(\lambda)$ lying
outside. This complex contour integral can be evaluated using calculus of
residues~\cite{Arfken2001} which becomes technically involved for large~$N$ and thus
limits the practical relevance of this route.

\subsubsection{Legendre transformation route:}
\label{sec:legendre-transf-rout}
The canonical Gibbs free energy
\begin{equation}
  \label{eq:11}
  \GFE(N)= - \kB T \ln \Q(N)
\end{equation}
and the grand potential
\begin{equation}
  \label{eq:12}
  \GFEGC(\mu) = - \kB T \ln \Z(\lambda)\eqspace,
\end{equation}
are related \via a Legendre transformation
\begin{equation}
  \label{eq:13}
  \GFE(N) = \GFEGC(\mu(N)) + N\trecdot\mu (N)\eqspace.
\end{equation}
The chemical potential~$\mu$ as a function of the segment number~$N$ is obtained by
inverting the relation
\begin{equation}
  \label{eq:14}
  N(\mu) =
  -\frac{\partial\GFEGC(\mu)}{\partial \mu}\eqspace.
\end{equation}
Let us briefly review the origin of \eqs~\eqref{eq:13} and~\eqref{eq:14} in the present
context.  Changing the integration variable in \eq~\eqref{eq:10} from $\lambda$ to $\mu =
\ln(\lambda)/\beta$, the complex path integral can be transformed into
\begin{equation}
  \label{eq:15}
  \Q(N) = \int_{\mathcal{C'}}\e^{-\beta \GFEGC(\mu) - \beta N \mu}\dd\mu
  \approx\e^{-\beta\GFEGC(\mu_{\mathrm{sp}}(N)) - \beta N\mu_{\mathrm{sp}}(N)}\eqspace,
\end{equation}
with the contour $\mathcal{C'} = \mu_0 + 2\pi\ci t/\beta$, $0\leq t\leq 1$ and $\mu_0=\kB
T\ln\lambda_0$. The integral in \eq~\eqref{eq:15} has been approximated by the method of
steepest descent, where the contour~$\mathcal{C'}$ is deformed such that it passes through
the saddle point~$\mu_{\mathrm{sp}}$~\cite{Arfken2001} determined by
equation~\eqref{eq:14}. If $\GFEGC$ features singularities deformation of the contour
$\mathcal{C'}$ requires extra care. In the present case the presence of a pole
$\lambdap=\exp(\beta\mu_{\mathrm{p}})$ of $\Z(\lambda)$ produces no problem as
$\mu_{\mathrm{sp}} < \mu_{\mathrm{p}}$ holds, meaning that the deformed contour does not
enclose the pole singularity.  This is different for the branch point
singularity~$\mu_{\mathrm{b}}$ where we will encounter the case
$\mu_{\mathrm{b}}<\mu_{\mathrm{sp}}$ for large $\loopexponent{} >2$.

\subsubsection{Dominating singularity route:}
\label{sec:domin-sing-route}
For large systems, \ie $N\gg1$, one approximately has $-\ln\Q(N)\sim N\ln\lambdad$, where
the dominant singularity $\lambdad=\exp(\beta\mu_{\mathrm{d}})$ is the singularity (in the
general case a pole or a branch point) of $\Z(\lambda)$ which has the smallest modulus.
One thus finds
\begin{equation}
  \label{eq:16}
  \GFE(N) = \kB T N \ln \lambdad\eqspace.
\end{equation}
This easily follows from \eq~\eqref{eq:10}: In the limit of $N\gg1$ the integral can be
approximated by expanding $\Z(\lambda)$ around $\lambdad$ and deforming $\mathcal{C}$ to a
Hankel contour which encircles~$\lambdad$~\cite{Flajolet1990}.  For the case where $N(\mu)
\propto (\mu_{\mathrm{d}} - \mu)^{-\alpha}$, $\alpha>0$, this can be understood also in
the context of a Legendre transform.  As $N=-\partial\GFEGC/\partial\mu$ one has
$\GFEGC\propto(\mu_{\mathrm{d}} - \mu)^{-\alpha+1}$ and therefore the first term of
\eq~\eqref{eq:13} scales like $\GFEGC(\mu(N))\propto N^{1-1/\alpha}$. Thus, the second
term $N\mu(N)\propto N \mu_{\mathrm{d}} - N^{1-1/\alpha} \propto N \mu_{\mathrm{d}}$ is
dominant.  Since the saddle point behaves as $\mu_{\mathrm{sp}} = \mu(N) \rightarrow
\mu_{\mathrm{d}}$ for $N\rightarrow\infty$, it follows that the dominating singularity
expression \eq~\eqref{eq:16} equals the Legendre transform \eq~\eqref{eq:13} in the
thermodynamic limit $N\rightarrow\infty$.

\subsection{Transfer matrix approach for $\mathbf{\loopexponent=0}$}
\label{sec:transf-matr-appr}

For $\loopexponent{}=0$ only interactions between nearest neighbors are present and
straight transfer matrix techniques are applicable.  We introduce a spin variable $i_n$
for each segment which can have the values $i_n=\mathrm{\Bstate,\Sstate,\Mstate}$. The
energetics are given by the Hamiltonian
\begin{equation}\label{eq:17}
  H(i_{1},i_{2},...,i_{N})%
  =\sum_{n=1}^{N}\g_{i_n}+\sum_{n=1}^{N-1}\V{i_{n}}{i_{n+1}}
\end{equation}
where $\g_{i_n}$ and $\V{i_{n}}{i_{n+1}}$ are the previously introduced parameters for the
segment and interfacial free energies.  The canonical partition function of the molecule
can be written as
\begin{equation}
  \label{eq:18}
  \Q(N) = \sum_{i_1, \ldots,i_N} \e^{-\beta H(i_{1},i_{2},...,i_{N}) }=
  \transpose{\vec v}
  \dotproduct
  \mat{T}^{N-1}\mat M_{\mathrm{TM}}
  \dotproduct
  \vec v
  \eqspace,
\end{equation}
where we introduced the transfer matrix $ \mat{T} = \mat{M}_{\mathrm{TM}}
\mat{V}_{\mathrm{TM}}$ and
\begin{equation}
  \label{eq:19}
  \mat M_{\mathrm{TM}}=
  \begin{pmatrix}\e^{-\beta\gB}&0&0\\0&\e^{-\beta\gS}&0\\0&0&\e^{-\beta\gM}\end{pmatrix}
  \eqspace,\quad
  \mat{V}_{\mathrm{TM}}=
  \begin{pmatrix}
    1&    \e^{-\beta\VBS}&    \e^{-\beta\VBM}\\
    \e^{-\beta\VSB}&    1 &    \e^{-\beta\VSM}\\
    \e^{-\beta\VMB}& \e^{-\beta\VMS}& 1
  \end{pmatrix}\eqspace,
  \quad
  \vec v =\begin{pmatrix}1\\1\\1\end{pmatrix}\eqspace.
\end{equation}
$\Q(N)$ is calculated readily by diagonalizing~$ \mat{T} $
\begin{equation}
  \label{eq:20}
  \Q(N) =   \transpose{\vec v} \dotproduct \mat U \mat D^{N-1}  \mat U^{-1} \mat M_{\mathrm{TM}} \dotproduct \vec v  =  
  \transpose{{\vec v}}_{\mathrm l}\dotproduct \mat D^{N-1}\dotproduct {\vec v}_{\mathrm r} 
  = \sum_{i=1}^3    v_{{\mathrm l},i}  v_{{\mathrm r},i} x_i^{N-1}
  \eqspace.
\end{equation}
where $\mat{D} = \mat U^{-1} \mat{T} \mat U$ is a diagonal matrix with eigenvalues $x_i$,
the columns of $\mat U$ are the right eigenvectors of~$ \mat{T} $, $\transpose{{\vec
    v}_{\mathrm l}} = \transpose{\vec v}\dotproduct \mat U$ and ${{\vec v}_{\mathrm
    r}}=\mat U^{-1}\mat M_{\mathrm{TM}} \dotproduct\vec v$.  By virtue of the
Perron-Frobenius theorem one eigenvalue $x_{\mathrm{max}}$ is larger than the other
eigenvalues and thus the free energy
is in the thermodynamic limit dominated by $x_{\mathrm{max}}$ and reads
\begin{equation}
  \label{eq:21b}
  \GFE= -\kB T \ln \Q(N)  \approx-\kB T N \ln x_{\mathrm{max}}\eqspace.
\end{equation}
The transfer-matrix eigenvalues $x_i$ and the poles $\lambdap$ of ${\Z}_{\loopexponent=0}$
\eq~\eqref{eq:7} are related \via $x_i = 1/\lambdap$.  As expected, the free energies from
the transfer matrix approach \eq~\eqref{eq:21b} and from the Poland-Scheraga approach for
$\loopexponent{}=0$ and using the dominating singularity approximation \eqs~\eqref{eq:16}
are identical in the limit $N\rightarrow \infty$.  Clearly, for $\loopexponent{}\neq 0$
the modified Poland-Scheraga approach yields new physics that deviates from the
transfer-matrix results.
 
Although not pursued in this paper, the transfer matrix approach allows to calculate
correlators. For example the probability~$p_{\Mstate{}}(k,m)$ of a denatured region with
$k$ consecutive molten base pairs starting at base $m$ 
is given by
\begin{equation}
  \label{eq:21}
  \begin{split}
    p_{\Mstate{}}(k,m)
    =\Q(N)^{-1}\trecdot\transpose{\vec v}\dotproduct \mat
    T^{m-2}(\mat{P}_{\mathrm{\Bstate}}\mat T + \mat{P}_{\mathrm{\Sstate}}\mat T)
    (\mat{P}_{\mathrm{\Mstate}} \mat T )^k (\mat{P}_{\mathrm{\Bstate}}\mat T +
    \mat{P}_{\mathrm{\Sstate}}\mat T) \mat T^{N-m-k-1}\mat M_{\mathrm{TM}}\dotproduct\vec
    v \eqspace.
  \end{split}
\end{equation}
The $\mat{P}_i$-matrices, which project a segment onto a certain state, are defined as
\begin{equation}
  \label{eq:22}
  \mat{P}_{\mathrm{\Bstate}}=\begin{pmatrix} 1&0&0\\0&0&0\\0&0&0\end{pmatrix}\eqspace,\quad
  \mat{P}_{\mathrm{\Sstate}}=\begin{pmatrix} 0&0&0\\0&1&0\\0&0&0\end{pmatrix}\eqspace,\text{ and} \quad
  \mat{P}_{\mathrm{\Mstate}}=\begin{pmatrix} 0&0&0\\0&0&0\\0&0&1\end{pmatrix}\eqspace.
\end{equation}

\section{Force extension curves}
\label{sec:force-extens-curve-1}

\ResubNew{The central quantity is $\GFE(\F,T,N)$, the Gibbs free energy of a DNA chain
  with $N$ base pairs, subject to a force~$\F$ and temperature~$T$.} From $\GFE(\F,T, N)$,
obtained via the Legendre transform, \eq~\eqref{eq:13}, the dominating singularity,
\eq~\eqref{eq:16}, or the exact transfer matrix partition function, \eq~\eqref{eq:20}, we
can calculate observables by performing appropriate derivatives.  The number of segments
in state $i=\mathrm{\Bstate,\Sstate,\Mstate}$ is obtained by
\begin{equation}
  \label{eq:23}
  N_i =  \left.\frac{\partial \GFE}{\partial \gi}\right|_{T,N,\F}\eqspace.
\end{equation}
The force extension curve is readily calculated \via
\begin{equation}
  \label{eq:24}
  x(\F)  = \left.-\frac{\partial \GFE}{\partial \F}\right|_{T,N} =
  - \sum_{i=\mathrm{\Bstate,\Sstate,\Mstate}}   \frac{\partial \GFE}{\partial
    \gi}\frac{\partial \gi}{\partial \F}
  = \sum_{i=\mathrm{\Bstate,\Sstate,\Mstate}} N_i\left(x^{\mathrm{WLC}}_i(\F) + 
    \F\DELETEcdot\segmentlength_i/\stretchmodulus_i\right)
\end{equation}
where $x_i^{\mathrm{WLC}}(\F)$ is the stretching response of a worm-like chain and given
explicitly in the supplementary information \eq~\eqref{appendix-eq:31}.

\subsection{Vanishing loop entropy, $\loopexponent{}= 0$}
\label{sec:affine-free-energies-1}

In this section we compare the prediction for vanishing loop exponent~$\loopexponent{}=0$
to experimental data and obtain estimates of the various parameters.  We also demonstrate
the equivalence of the grand canonical and canonical ensembles even for small chain
length~$N$.
\label{sec:parameters}

In order to reduce the number of free fitting parameters we extract as many reasonable
values from literature as possible.  For the helical rise, the stretch modulus, and
persistence length of \Bstate-DNA we use $\segmentlengthB=\unit{3.4}{\angstrom}$,
$\stretchmodulusB=\unit{1}{\nano\newton}$, and
$\persistencelengthB=\unit{48}{\nano\meter}$~\cite{Wenner2002}. \ResubNew{For the
  \Mstate-state, which is essentially single stranded DNA (ssDNA), \emph{ab initio}
  calculations yield $\segmentlengthM=\unit{7.1}{\angstrom}$ and
  $\stretchmodulusM=\unit{2\trecdot9.4}{\nano\newton}$~\cite{Hugel2005}, where
  $\stretchmodulusM$ is valid for small forces $\F<\unit{400}{\pico\newton}$ and the
  factor~$2$ accounts for the presence of two ssDNA strands.  Our value for the stretch
  modulus is considerably larger than previous experimental fit
  estimates~\cite{Smith1996,Dessinges2002} which might be related to the fact that
  experimental estimates depend crucially on the model used to account for conformational
  fluctuation effects; however, the actual value of $\stretchmodulusM$ is of minor
  importance for the stretching response, see supporting material.  The persistence length
  of ssDNA is given by
  $\persistencelengthM\approx\unit{3}{\nano\meter}$~\cite{Murphy2004}.  It turns out that
  the quality of the fit as well as the values of the other fit parameters are not very
  sensitive to the exact value of the persistence length~$\persistencelengthS$ and the
  stretch modulus~$\stretchmodulusS$ of the \Sstate-state as long as
  $\unit{10}{\nano\meter}\lesssim\persistencelengthS\lesssim\unit{50}{\nano\meter}$ and
  $\stretchmodulusS$ is of the order of $\stretchmodulusM$, see supporting material
  section~\ref{appendix-sec:effect-param-feat}}. Therefore we set
$\persistencelengthS=\unit{25}{\nano\meter}$, which is an intermediate value between the
persistence lengths of ssDNA and \Bstate-DNA, and $\stretchmodulusS = \stretchmodulusM =
\unit{2\trecdot9.4}{\nano\newton}$~\cite{Cocco2004}.  The segment length of the
\Sstate-state $\segmentlengthS$ will be a fit parameter.

The chemical potentials $\gio$, $i=\mathrm{\Bstate,\Sstate,\Mstate}$, account for the free
energy of base pairing and, since we set the interaction energies between neighboring
segments of the same type to zero, $\Vii = 0$, also for the free energy gain due to base
pair stacking~\cite{SantaLucia1998}.  They also correct for the fact that the reference
state of the three different WLCs, which is $x = 0$ in the Helmholtz ensemble (constant
extension $x$), \cf \eq~\eqref{appendix-eq:30}, is not the same as contour and persistence
lengths differ for \Bstate-, \Sstate-, and molten \Mstate-DNA. We choose $\gBo=0$ and
treat $\gSo$ and $\gMo$ as fitting parameters. 

\ResubNew{Each of these parameters controls a distinct feature of the force-extension
  curve: The chemical potentials~$\gio$ determine the critical forces, the segment
  lengths~$\segmentlength_i$ affect the maximal extensions of each state and the
  off-diagonal $\Vij$ control the cooperativity of the transitions, see supporting
  material section~\ref{appendix-sec:effect-param-feat} for an illustration and
  section~\ref{appendix-sec:tables-with-model} for a summary of all parameter values.}

\subsubsection{Force extension curve}
\label{sec:force-extens-curve-2}
\begin{figure}
  \centering
  \includegraphics{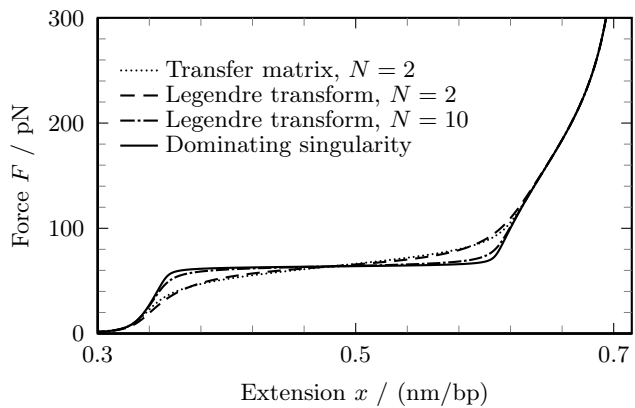}
  \caption{Comparison of force extension curves obtained by different methods for
    $\loopexponent{}=0$.  The curve obtained \via the exact transfer matrix calculation
    \eq~\eqref{eq:20} is already for $N=2$ accurately reproduced by the approximate
    Legendre transformation \eq~\eqref{eq:13}.  The dominating singularity method
    \eqs~\eqref{eq:16} or -- equivalently -- \eqref{eq:21b} is strictly valid in the
    thermodynamic limit but agrees with the Legendre transform already for a modest
    value of $N=10$. \ResubNew{The units of the abscissa is extension per base
      pair~(bp).}  Parameters for $\lambda$-DNA in the absence of DDP are
    used, see supporting material section~\ref{appendix-sec:tables-with-model}.}
  \label{fig:1}
\end{figure}

In \fig~\ref{fig:1} force extension curves based on three different levels of
approximation are compared, using the same parameters that we extracted from DNA
stretching data as will be detailed below.  It turns out that the force extension curve
obtained \via the Legendre transformation route \eq~\eqref{eq:13} (dashed line) is a very
good approximation of the results obtained from the exact transfer matrix results
\eq~\eqref{eq:20} (dotted line) already for $N=2$.  For $N=10$ and larger virtually no
differences between these two approaches are detectable.  The deviations from the
dominating singularity route \eq~\eqref{eq:16} (solid line), which gives a result
independent of $N$, are somewhat larger.  But one sees that already the Legendre transform
for $N=10$ (dash-dotted line) matches the dominating singularity result very closely.
Therefore the use of the dominating singularity, \eq~\eqref{eq:16}, or the largest
transfer-matrix eigenvalue, \eq~\eqref{eq:21b}, is a very good approximation already for
oligo-nucleotides and will be used in the rest of this work.

\begin{figure}
  \centering
  \includegraphics{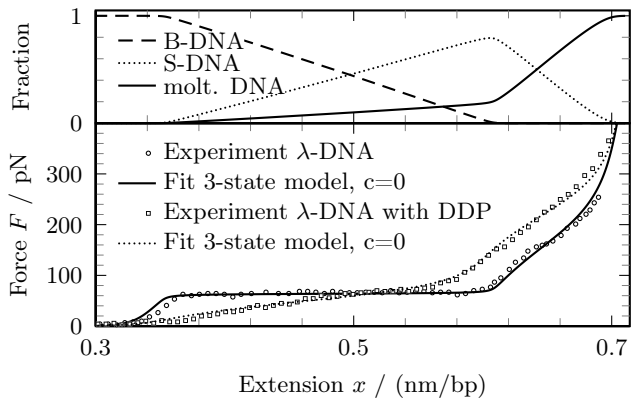}
  \caption{Bottom panel: Force extension curve of double-stranded $\lambda$-DNA with and
    without DDP. Symbols denote experimental data~\cite{Krautbauer2000}, lines are fits
    with the three-state model for $\loopexponent{}=0$. The main difference between the
    two curves is the lack of cooperativity in the \Bstate\Sstate-transition in the
    presence of DDP which we take into account by choosing vanishing interaction
    energies $\Vij=0$, $i,j=\mathrm{\Bstate,\Sstate,\Mstate}$.  Top panel: Fraction
    $N_i/N$ of segments in the different states as follows from \eq~\eqref{eq:23} in the
    absence of DDP. }  \label{fig:2}
\end{figure}

In \fig~\ref{fig:2} experimental stretching curves of $\lambda$-DNA with and without DDP
(cisplatin) are presented~\cite{Krautbauer2000}.  When \Bstate-DNA is converted into
\Sstate-DNA or \Mstate-DNA the base stacking is interrupted which gives rise to an
interfacial energy between \Bstate{} and \Sstate{} as well as between \Bstate{} and
\Mstate{} of the order of the stacking energy~\cite{Konrad1996,Kosikov1999}.
\ResubNew{For untreated DNA, we use the value $\VBS=\VSB=\VBM=\VMB=
  \unit{1.2\trecdot10^{-20}}{\joule}$ and show in the supplement that variations down to
  $\unit{0.8\trecdot10^{-20}}{\joule}$ do not change the resulting curves much. $\VSM$ is
  presumably small as the stabilizing stacking interactions are already
  disrupted~\cite{Cocco2004}, we thus set $\VSM=\VMS=0$ for the fits in \fig~\ref{fig:2}
  --~but we will come back to the issue of a non-zero $\VSM$ later on.}  Cisplatin is
thought to disrupt the stacking interaction between successive base pairs and thereby to
reduce the cooperativity of the \Bstate\Sstate-transition. This fact we incorporate by
setting all interfacial energies to zero, $\Vij = 0$, for DDP treated DNA.  The three
remaining undetermined parameters ($\segmentlengthS$, $\gSo$, $\gMo$) have distinct
consequences on the force-extension curve.  The segment length $\segmentlengthS$ and the
chemical potential $\gSo$ determine the position of the \Bstate{}\Sstate{}-plateau with
respect to the polymer extension and applied force, respectively, while the chemical
potential $\gMo$ controls the force at which the second transition appears.  Fitting to
experimental data is thus straightforward and yields for untreated $\lambda$-DNA
$\segmentlengthS = \unit{6.1}{\angstrom}$, $\gSo = \unit{1.6\trecdot10^{-20}}{\joule}$,
$\gMo = \unit{2.4\trecdot10^{-20}}{\joule}$ and for $\lambda$-DNA in the presence of DDP
(cisplatin) $\segmentlengthS = \unit{6.0}{\angstrom}$, $\gSo =
\unit{1.2\trecdot10^{-20}}{\joule}$, $\gMo = \unit{2.8\trecdot10^{-20}}{\joule}$, see
\fig~\ref{fig:2}.  We also fit the number of monomers~$N$ and allow for an overall shift
along the $x$-axis.  \ResubNew{The main difference between the two stretching curves is
  the cooperativity of the \Bstate\Sstate-transition, which is controlled by the
  interfacial energies $\VBS$ and $\VBM$. Note that, although the over-stretching
  transition is quite sharp for DNA without DDP, it is not a phase transition in the
  strict statistical mechanics sense.  A true phase transition arises only for
  $\loopexponent{}>1$, as will be shown in the next section.} The fitted value of $\gMo$
is about two times larger than typical binding energies~\cite{SantaLucia1998} for pure
DNA.  \ResubNew{As a possible explanation, we note that the force extension curve of DNA
  without DDP exhibits pronounced hysteresis (especially at higher force) which will
  increase the apparent binding energy due to dissipation effects~\cite{Ho2009}.  Any
  statements as to the stability of \Sstate-DNA based on our fitting procedures are thus
  tentative.  However, such complications are apparently absent in the presence of
  DDP~\cite{Krautbauer2000} which rules out kinetic effects as the reason for our
  relatively high fit values of $\gMo$ and the stability of \Sstate-DNA. Cisplatin most
  likely stabilizes base pairs due to cross-linking and thus shifts the subtle balance
  between \Bstate{}-, \Sstate{}-, and \Mstate{}-DNA. Therefore, the relative stability of
  \Bstate{}-, \Sstate{}-, and \Mstate{}-DNA is sensitively influenced by co-solute
  effects.  We note that even with $\loopexponent{}=0$ a good fit of the data is possible.
  In the top panel of \fig~\ref{fig:2} we show the fraction of segments in \Bstate{}-,
  \Sstate{}-, and \Mstate{}-states for untreated $\lambda$-DNA. There is a balanced
  distribution of bases in all three states across the full force range, in agreement with
  previous results~\cite{Cocco2004}.}

\subsection{Non-vanishing loop entropy, $\loopexponent{}\neq0$}
\label{sec:logar-loop-free}

\begin{figure}
  \centering
  \includegraphics{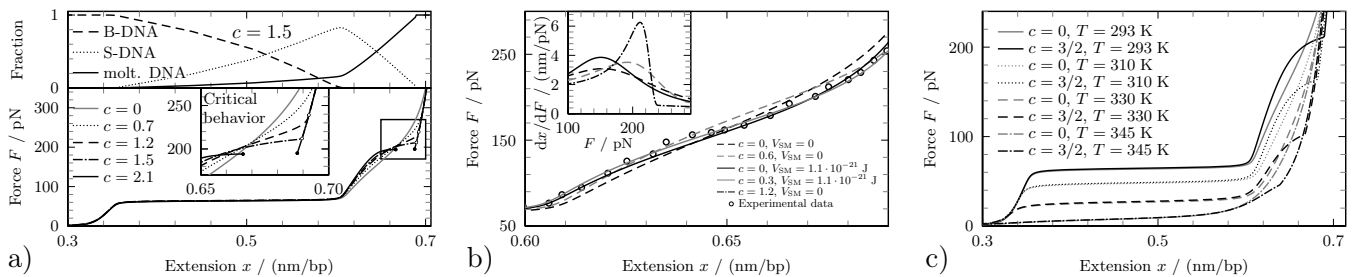}
  \caption{ Various force-extension curves of the three-state model with fit parameters
    for $\lambda$-DNA without DDP.  \textbf{a)} Lower panel: Force extension curves for
    different values of the loop exponent~$\loopexponent$, showing no phase transition
    ($\loopexponent\leq1$), a continuous ($1<\loopexponent\leq2$), or a first order
    phase transition ($\loopexponent>2$). The critical forces are denoted by open
    circles. The inset is a magnification of the region around the transition.  Upper
    panel: Fraction of bases in the three states for $\loopexponent{}=3/2$. The critical
    transition, above which all bases are in the molten \Mstate-state, is discerned as a
    kink in the curves.  \textbf{b)} Comparison of experimental data (circles) and
    theory for $\loopexponent{}\neq 0$.  The curve for $\loopexponent{}=0$ and $\VSM=0$,
    already shown in \fig~\ref{fig:2}, is obtained by fitting $\segmentlengthS$, $\gSo$,
    $\gMo$ to the experimental data, the values of which are kept fixed for all curves
    shown.  The curve for $\VSM=0$ and $\loopexponent{}=0.6$ results by fitting
    $\loopexponent{}$ and slightly improves the fit quality.  The curve
    $\loopexponent{}=0$ and $\VSM = \unit{1.1\trecdot10^{-21}}{\joule}$ is obtained by
    fitting $\VSM$.  The curve for $\VSM = \unit{1.1\trecdot10^{-21}}{\joule}$ and
    $\loopexponent{}=0.3$ is obtained by fitting $\loopexponent{}$ and keeping $\VSM$
    fixed.  The inset shows the first derivative of $x(\F)$ and illustrates that
    increasing $\loopexponent{}$ leads to a growing asymmetry around the transition
    region.  \textbf{c)} Temperature dependence of the force extension curves.
    Increasing temperature leads to a decrease of the \Bstate\Sstate-plateau force.  In
    the presence of a true denaturing transition, i.e. for $\loopexponent{}>1$, the
    critical force~$\Fc$ decreases with increasing temperature and for $\F>\Fc$ the
    force extension curve follows a pure WLC behavior. }
  \label{fig:3}
\end{figure}

We now turn to non-zero loop exponents $\loopexponent{} \neq 0$ and in specific try to
estimate $\loopexponent{}$ from the experimental stretching data.  The partition
function~$\Z_{\loopexponent \neq 0}$ in \eq~\eqref{eq:9} exhibits two types of
singularities. First, simple poles at $\lambda=\lambdap$, which are the zeros of the
denominator of \eq~\eqref{eq:9} and which are determined as the roots of the equation
\begin{equation}
  \label{eq:25}
  -\frac{b_5 + b_6 \lambda + b_7\lambda^2}{b_8 \lambda + b_9\lambda^2} =  \Li{\loopexponent}{\lambda /\lambdab}
  \eqspace.
\end{equation}
Second, a branch point that occurs at
\begin{equation}
  \label{eq:26}
  \lambda = \lambdab = \e^{\beta \gM}
  \eqspace.
\end{equation}
The singularity with the smallest modulus is the dominant
one~\cite{Poland1966,Flajolet1990}, and we define the critical force~$\Fc$ as the force
where both equations, \eq~\eqref{eq:25} and~\eqref{eq:26}, hold simultaneously.  For
$\loopexponent\leq1$ the dominant singularity is always given by the pole~$\lambdap$ and
thus no phase transition is possible. For $1<\loopexponent\leq2$ a continuous phase
transition occurs. By expanding \eq~\eqref{eq:25} around $\Fc$ one can show that all
derivatives of the free energy up to order~$n$ are continuous, where $n\in\Nbb$ is defined
as the largest integer with $n<(\loopexponent-1)^{-1}$~\cite{Kafri2002}. For instance,
$\loopexponent=3/2$ leads to a kink in the force extension curve. For $\loopexponent=1.2$
this leads to a kink in $x'''(\F)$. If $\loopexponent>2$ the transition becomes first
order and the force extension curve exhibits a discontinuity at $\F=\Fc$. In
\fig~\ref{fig:3}a we plot force extension curves for different values of the loop
exponent~$\loopexponent$ with all other parameter fixed at the values fitted for untreated
DNA.  It is seen that finite $\loopexponent{}$ leads to changes of the force extension
curves only at rather elevated forces. In order to see whether a finite $\loopexponent{}$
improves the comparison with the experiment and whether it is possible to extract the
value of $\loopexponent{}$ from the data, we in \fig~\ref{fig:3}b compare the untreated
DNA data with a few different model calculations for which we keep the parameters
$\segmentlengthS$, $\gSo$, $\gMo$, $\VBS$, $\VBM$ fixed at the values used for the fit
with $\loopexponent{}=0$ in \fig~\ref{fig:2}. Allowing for finite $\loopexponent{}$ but
fixing a zero domain wall energy between \Sstate- and \Mstate-regions, $\VSM=0$, leads to
an optimal exponent $\loopexponent{}=0.6$ and slightly improves the fit to the data which
show the onset of a plateau at a force of about $\unit{100}{\pico\newton}$.  The same
effect, however, can be produced by fixing $\loopexponent{}=0$ and allowing for a finite
$\VSM$, which yields the optimal value of $\VSM = \unit{1.1\trecdot10^{-21}}{\joule}$.
Finally, fixing $\VSM = \unit{1.1\trecdot10^{-21}}{\joule}$ and optimizing
$\loopexponent{}$ yields in this case $\loopexponent{}=0.3$ and perfect agreement with the
experimental data. However, the significance of this improvement is not high, as the
experimental data are quite noisy and possibly plagued by kinetic effects.  What the
various curves illustrate quite clearly, however, is that a non-zero exponent
$\loopexponent{}$ leads to modifications of the stretching curves that are similar to the
effects of a non-vanishing domain wall energy $\VSM$. Although $\VSM$ should be
considerably smaller than $\VBM$ or $\VBS$, a finite value of $\VSM =
\unit{1.1\trecdot10^{-21}}{\joule}$ as found in the fit is reasonable and cannot be ruled
out on general grounds.  The maximal value of $\loopexponent{}$ is obtained for vanishing
$\VSM$ and amounts to about $\loopexponent{}=0.6$.  A value of $\loopexponent{}=2.1$,
which would be expected based on the entropy of internal DNA loops~\cite{Kafri2002}, on
the other hand does not seem compatible with the experimental data, as follows from
\fig~\ref{fig:3}a. This might have to do with the presence of nicks.  \ResubNew{Nicks in
  the DNA drastically change the topology of loops and result in a reduced loop exponent
  which is $\openendexponent{}=0$ for an ideal polymer and $0.092$ for a self avoiding
  polymer~\cite{Kafri2002}. Therefore, the low value of $\loopexponent{}$ we extract from
  experimental data might be a signature of nicked DNA.  Additional effects such as salt
  or co-solute binding to loops are also important. Therefore $\loopexponent$ can be
  viewed as a heuristic parameter accounting for such non-universal effects as well.  We
  note in passing that $\loopexponent$ only slightly affects the
  \Bstate\Sstate-transition, as seen in \fig~\ref{fig:3}a.}

\section{Finite temperature effects}
\label{sec:phase-diagram}

The temperature dependence of all parameters is chosen such that the force extension
curves at $T=\unit{20}{\degreecelsius}$ that were discussed up to now remain unchanged.
The persistence lengths are modeled as
$\persistencelength_i(T)=\persistencelength_i\trecdot(T/\unit{293}{\kelvin})^{-1}$.  The
\Sstate-state free energy is split into enthalpic and entropic parts as $\gSo(T)
=\scalegSo (h_{\mathrm{\Sstate}} - Ts_{\mathrm{\Sstate}})$ where we use
$h_{\mathrm{\Sstate}}=\unit{7.14\trecdot10^{-20}}{\joule}$ and
$s_{\mathrm{\Sstate}}=\unit{1.88\trecdot10^{-22}}{\joule\per\kelvin}$
from~\citet{ClausenSchaumann2000}. The correction factor $\scalegSo = 0.98$ accounts for
slight differences in the experimental setups and is determined such that
$\gSo(T=\unit{20}{\degreecelsius})$ equals the previously fitted value.  The molten state
energy $\gMo(T)=(h_{\mathrm{\Mstate}} - Ts_{\mathrm{\Mstate}})$ is also chosen such that
$\gMo(T=\unit{20}{\degreecelsius})$ agrees with the previous fit value and that the
resulting denaturing temperature in the absence of force agrees with experimental
data. Assuming a melting temperature of $\Tc = \unit{348}{\kelvin}$ for
$\lambda$-DNA~\cite{Gotoh1976}, we obtain $\gMo(T) =
\unit{1.5\trecdot10^{-19}}{\joule}-T\trecdot\unit{4.2\trecdot10^{-22}}{\joule\per\kelvin}$
for $\loopexponent{} = 0$ and $\gMo(T) =
\unit{1.6\trecdot10^{-19}}{\joule}-T\trecdot\unit{4.6\trecdot10^{-22}}{\joule\per\kelvin}$
for $\loopexponent{} = 3/2$.  In \fig~\ref{fig:3}c we plot a few representative stretching
curves for different temperatures.  It is seen that increasing temperature lowers the
\Bstate\Sstate-plateau and makes this transition less cooperative. Differences between
$\loopexponent{}=0$ and $\loopexponent{}=3/2$ are only observed at elevated forces, where
for $\loopexponent{}=3/2$ one encounters a singularity characterized by a kink in the
extension curves.

\begin{figure}
  \centering
  \includegraphics{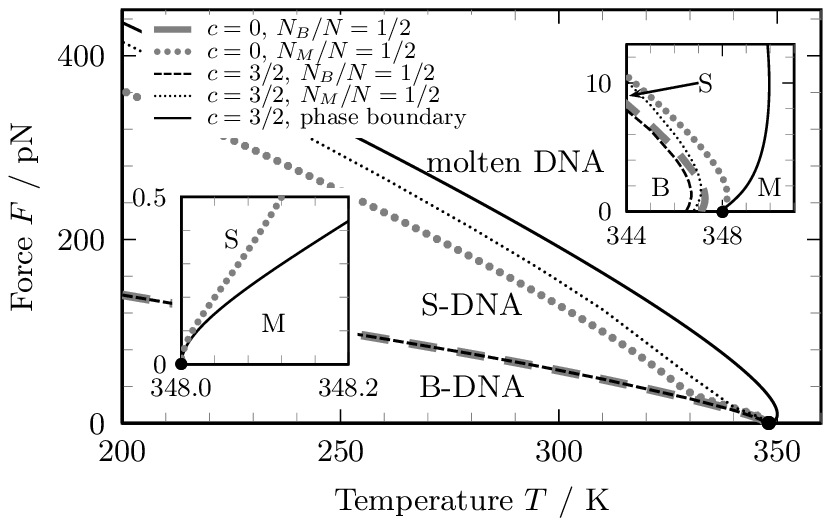}
  \caption{ \ResubNew{The solid line shows the critical force $\Fc$ for
      $\loopexponent{}=3/2$, at which a singularity occurs according to
      \eq~\eqref{eq:27}.  Phase boundaries for $\loopexponent{}=0$ (thick lines) and
      $\loopexponent{}=3/2$ (thin lines) are defined by $N_{\mathrm{\Mstate}}/N=0.5$
      (dotted) and $N_{\mathrm{\Bstate}}/N=0.5$ (dashed).  The melting temperature~$\Tc$
      is denoted by the dot. The insets show the behavior of the phase boundaries near
      the melting temperature, $\F \propto \sqrt{\Tc-T}$.}  Parameters for $\lambda$-DNA
    without DDP are used.  }
  \label{fig:4}
\end{figure}

For $\loopexponent{}>1$ the critical force~$\Fc$ is defined as the force where the pole
and the branch point coincide and \eqs~\eqref{eq:25} and~\eqref{eq:26} are simultaneously
satisfied.  The phase boundary in the force-temperature plane is thus defined by
\begin{equation}
  \label{eq:27}
  -\frac{b_5 + b_6 \lambdab + b_7\lambdab^2}{b_8 \lambdab + b_9\lambdab^2}=\zeta(\loopexponent)
  \eqspace,
\end{equation}
where $b_i$ and $\lambdab$ depend on $T$ and $\F$ and $\zeta(\loopexponent{}) =
\Li{\loopexponent}{1}$ is the Riemann zeta function. The phase boundary $\Fc(T)$ for
$\loopexponent{}=3/2$ is shown in \fig~\ref{fig:4} and agrees qualitatively with the
experimental data of ref.~\cite{Williams2001a}.  For exponents $\loopexponent{}<1$ no true
phase transition exists.  We therefore define crossover forces as the force at which half
of the segments are in the molten state or in the \Bstate-state,
i.e. $N_{\mathrm{\Mstate}}/N=1/2$ or $N_{\mathrm{\Bstate}}/N=1/2$.  In \fig~\ref{fig:4} we
show these lines for both $\loopexponent{}=0$ and $\loopexponent{}=3/2$. Note that the
parameters for the $\loopexponent{}=0$ case have been adjusted so that
$N_{\mathrm{\Mstate}}/N=1/2$ at $T=\unit{348}{\kelvin}$ and $\F=0$.  The broken lines on
which $N_{\mathrm{\Bstate}}/N=1/2$ for $\loopexponent{}=0$ and $\loopexponent{}=3/2$ are
virtually the same, showing again that loop entropy is irrelevant for the
\Bstate\Sstate-transition.  The \Sstate-state is populated in the area between the
$N_{\mathrm{\Bstate}}/N=1/2$ and $N_{\mathrm{\Mstate}}/N=1/2$ lines, which for
$T>\unit{330}{\kelvin}$ almost coalesce, meaning that at elevated temperatures the
\Sstate-state is largely irrelevant.  Force-induced re-entrance at constant temperature is
found in agreement with previous two-state models~\cite{Hanke2008,Rahi2008a}.  Re-entrance
at constant force as found for a Gaussian model~\cite{Hanke2008} is not reproduced, in
agreement with results for a non-extensible chain~\cite{Rahi2008a}.

\begin{figure}
  \centering
  \includegraphics{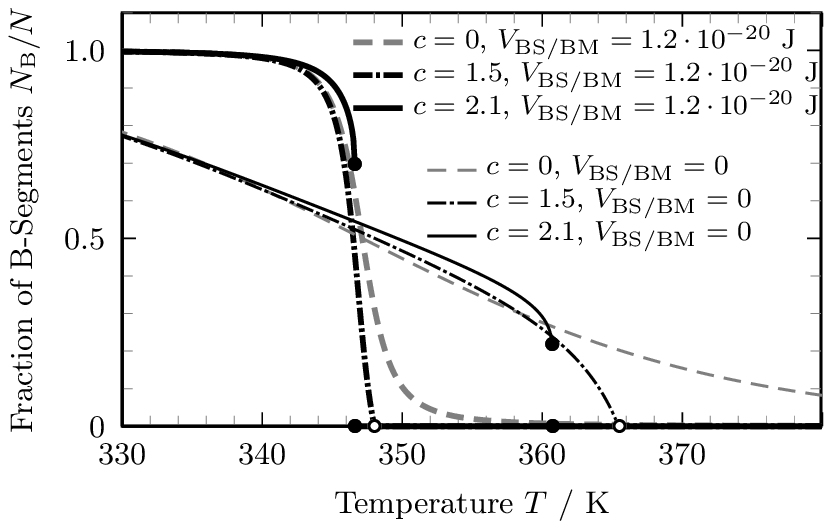}
  \caption{Relative fraction of segments in the \Bstate-state, $N_{\mathrm\Bstate}/N$,
    as a function of temperature for different loop exponents $\loopexponent=0,\ 1.5,\
    2.1$ and for finite \Bstate\Sstate{} interfacial energy
    $\VBS=\VBM=\unit{1.2\trecdot10^{-20}}{\joule}$ (bold lines) and for $\VBS=\VBM=0$
    (thin lines).  Circles indicate the positions of the phase transition. For all
    curves parameters for $\lambda$-DNA without DDP have been used, $\gMo(T) =
    \unit{1.5\trecdot10^{-19}}{\joule}-T\trecdot\unit{4.2\trecdot10^{-22}}{\joule\per\kelvin}$
    for $\loopexponent{} = 0$ and $\gMo(T) =
    \unit{1.6\trecdot10^{-19}}{\joule}-T\trecdot\unit{4.6\trecdot10^{-22}}{\joule\per\kelvin}$
    for $\loopexponent{} > 0$.  }
  \label{fig:5}
\end{figure}

As we have shown so far, a non-zero loop exponent~$\loopexponent{}$ only slightly improves
the fit of the experimental stretching data and the optimal value found is less than
unity.  This at first sight seems at conflict with recent theoretical work that argued
that a loop exponent larger than $\loopexponent{}=2$ is needed in order to produce
denaturation curves (at zero force) that resemble experimental curves in terms of the
steepness or cooperativity of the transition~\cite{Kafri2002}.  To look into this issue,
we plot in \fig~\ref{fig:5} the fraction of native base pairs, $N_{\mathrm{\Bstate}}/N$ as
a function of temperature for zero force and different parameters. As soon as the
domain-wall energies $\VBS$ and $\VBM$ are finite, the transition is quite abrupt, even
for vanishing exponent $\loopexponent{}$.  Therefore even loop exponents $\loopexponent<1$
yield melting curves which are consistent with experimental data, where melting occurs
over a range of the order of $\unit{10}{\kelvin}$~\cite{Blossey2003,Gotoh1976}.

\section{Conclusions}
\label{sec:conclusions}

The fact that the domain-wall energy due to the disruption of base pair stacking, $\VBS$
and $\VBM$, and the loop entropy embodied in the exponent $\loopexponent{}$, give rise to
similar trends and sharpen the melting transition has been realized and discussed
before~\cite{Blossey2003,Whitelam2008}.  The present three-state model and the
simultaneous description of experimental data where the denaturation is induced by
application of force and by temperature allows to disentangle the influence of these two
important effects.  By the application of a force, the de-stacking and the loop formation
occur subsequently, allowing to fit both parameters separately. As our main finding, we
see that for a finite domain-wall energy $\VBS=\VBM$, the additional influence of the loop
exponent on the force stretching curves and the denaturation curves is small. In fact, the
optimal value for $\loopexponent{}$ turns out to be of the order of $\loopexponent{}
\approx 0.3 - 0.6$, even if we choose a vanishing value $\VSM=0$.  This estimate for
$\loopexponent{}$ is smaller than previous estimates.  One reason for this might be nicks
in the DNA. So it would be highly desirable to redo stretching experiments with un-nicked
DNA \ResubNew{from which the value of $\loopexponent{}$ under tension could be determined.
  A second transition at high forces of about $\F \simeq \unit{200}{\pico\newton}$ which
  is seen in the experimental data used in the paper, inevitably leads via the fitting
  within our three-state model to an intermediate \Sstate{}-DNA state. But we stress that
  the occurrence of such an intermediate \Sstate{}-state depends on the fine-tuning of all
  model parameters involved, which suggests that in experiments the \Sstate{}-state
  stability should also sensitively depend on the precise conditions.}  One drawback of
the current model is that sequence effects are not taken into account.  This means that
our fitted parameters have to be interpreted as coarse-grained quantities which average
over sequence disorder.  Calculations with explicit sequences have been done for short DNA
strands but should in the future be doable for longer DNA as well.

\section{Acknowledgments}
\label{sec:acknowledgements}

The authors express their gratitude to Hauke Clausen-Schaumann and Rupert Krautbauer for
providing the experimental data and helpful comments and thank Ralf Metzler, Matthias
Erdmann, and Dominik Ho for fruitful discussions.  Financial support comes from the DFG
\via grants NE 810/7 and SFB 863.  T.R.E. acknowledges support from the
\emph{Elitenetzwerk Bayern} within the framework of \emph{CompInt}.

\bibliography{dna_BSL-bib}

\end{document}


\title{Supporting material\\A three-state model with loop entropy for the over-stretching
  transition of DNA}

\author{Thomas R. Einert} \email[E-mail: ]{einert@ph.tum.de} \affiliation{Physik
  Department, Technische Universit\"at M\"unchen, 85748 Garching, Germany}%
\author{Douglas B. Staple} \affiliation{Department of Physics and Atmospheric Science,
  Dalhousie University, Halifax, B3H 3J5 Canada}\affiliation{Max-Planck-Institut f\"ur
  Physik komplexer Systeme, 01187 Dresden, Germany}%
\author{Hans-J\"urgen Kreuzer} \affiliation{Department of Physics and Atmospheric Science,
  Dalhousie University, Halifax, B3H 3J5 Canada}%
\author{Roland R. Netz} \email[E-mail: ]{netz@ph.tum.de} \affiliation{Physik Department,
  Technische Universit\"at M\"unchen, 85748 Garching, Germany}

\date{\today}
\begin{abstract}
  The supporting material shows the calculation of the of Gibbs free energy $\giWLC(\F)$
  of one worm-like chain segment subject to a force~$\F$. Additionally the explicit form
  of the grand canonical partition function~$\Z$ is given. The loop exponent of an ideal
  polymer is derived and effects of self-avoidance on the loop exponent are discussed. We
  illustrate that different model parameters are linked to certain features of the force
  extension curve. Tables with all parameters used in the main paper are given at the end
  of this document.
\end{abstract}
\keywords{DNA, denaturation, over-stretching, phase transition}

\maketitle
\tableofcontents
\section{Gibbs free energy of a worm-like chain}\label{sec:gibbs-free-energy}

To calculate the worm-like chain Gibbs free energy $\giWLC(\F)$, we first calculate the
Helmholtz free energy $\HFEWLC_i(x)$ of a worm-like chain (WLC) in the fixed extension
ensemble (Helmholtz ensemble). After that we Legendre transform $\HFEWLC_i (x)$ in order
to obtain $\GFEWLC_i(\F)$, which is the thermodynamic potential of a WLC in the fixed
force ensemble (Gibbs ensemble). As we will see, $\GFEWLC_i(\F)$ is extensive with respect
to the chain's contour length $n \segmentlength_i$, $\segmentlength_i$ is the length of
one segment, and thus we can define the Gibbs free energy per segment $\giWLC(\F) =
\GFEWLC_i(\F) / n$.  We introduce the dimensionless variables
\begin{align}
  \label{eq:28}
  \tilde x_i &= \frac{x}{n \segmentlength_i}\eqspace, &\tilde \F_i &= \F \cdot
  \beta\persistencelength_i\eqspace, & \HFEWLCdl_i &= \HFEWLC_i \cdot \frac{\beta
    \persistencelength_i}{n \segmentlength_i}\eqspace, &\GFEWLCdl_i &= \GFEWLC_i
  \cdot\frac{\beta \persistencelength_i}{n \segmentlength_i}\eqspace,
  \eqcomment{$i=\mathrm{\Bstate,\Sstate,\Mstate}$,}
\end{align}
where $\persistencelength_i$ is the persistence length of DNA in state
$i=\mathrm{\Bstate,\Sstate,\Mstate}$.  The interpolation formula of \citet{Marko1995} for
the force extension curve of the WLC
\begin{equation}
  \label{eq:29}
  \tilde \F_i(\tilde x_i) = \frac{1}{4(1-\tilde x_i)^2} + \tilde x_i - \frac14
\end{equation}
is used to calculate the free energy in the Helmholtz ensemble
\begin{equation}
  \label{eq:30}
  \HFEWLCdl_i(\tilde x) = \int_0^{\tilde x} \tilde  \F(\tilde x') \dd \tilde x' + \HFEWLCdl_{i,0}= \frac{\tilde
    x^2 (2\tilde x - 3)}{4 (\tilde x -1 )} + \HFEWLCdl_{i,0}\eqspace.
\end{equation}
$\HFEWLCdl_{i,0}$ is a free energy offset that accounts for the fact that even in the
absence of an external force the free energies of chains consisting of \Bstate-, \Sstate-,
or \Mstate{}-segments are not the same.  In fact, this constant is not easy to calculate
but can be dropped in the following since it can be adsorbed into the free energy
contributions $\gio$, \cf \eq~\eqref{main-eq:2} in the main text.

Now, we invert \eq~\eqref{eq:29}
\begin{equation}
  \label{eq:31}
  \tilde x_i (\tilde\F_i) = 1 - \left[
    \frac{
      \frac{4{\tilde\F_i}}{3}-1
    }{
      \left(2+\sqrt{4-\left(\frac{4{\tilde\F_i}}{3}-1\right)^3}\right)^{1/3}
    }
    + \left(2 + \sqrt{4 - \left(\frac{4{\tilde \F_i}}{3}-1\right)^3}\right)^{1/3}
  \right]^{-1}
\end{equation}
and perform a Legendre transformation to obtain the Gibbs free energy of a WLC in the
Gibbs ensemble
\begin{equation}
  \label{eq:32}
  \GFEWLCdl_i(\tilde\F_i)= \HFEWLCdl_i(\tilde x_i(\tilde\F_i))
  -\tilde\F_i \cdot \tilde x_i(\tilde \F_i)\eqspace.
\end{equation}
\Eqs~\eqref{eq:28} and~\eqref{eq:32} lead to
\begin{equation}
  \label{eq:33}
  \giWLC(\F)=\frac1n\cdot\frac{n \segmentlength_i}{\beta \persistencelength_i}\cdot\GFEWLCdl(\F\beta\persistencelength_i) =  \frac{\segmentlength_i}{\beta\persistencelength_i}\left(
    \frac{
      \tilde x_i\left({\F}{\beta\persistencelength_i}\right)^2
      (2\tilde x_i\left({\F}{\beta\persistencelength_i}\right) - 3)
    }{
      4 (\tilde x_i\left({\F}{\beta\persistencelength_i}\right) -1 )
    }
    -  {\F}{\beta\persistencelength_i}\cdot
    \tilde x_i\left({\F}{\beta\persistencelength_i}\right)
  \right)\eqspace,
\end{equation}
with $\tilde x_i(\tilde \F_i)$ from \eq~\eqref{eq:31}.

\section{Explicit form of the grand canonical partition function}
\label{sec:near-neighb-inter}

In the main text the expression
\begin{equation}
  \label{eq:34}
  \Z = \sum_{k=0}^\infty\transpose{\vec v}\cdot(\mat M_{\mathrm{PS}} \mat V_{\mathrm{PS}})^k\mat M_{\mathrm{PS}}\cdot\vec v
  =\transpose{\vec v}\cdot(\mat{1} - \mat M_{\mathrm{PS}}\mat V_{\mathrm{PS}})^{-1}\mat M_{\mathrm{PS}}\cdot\vec v\eqspace,
\end{equation}
has been derived for the grand canonical partition function. The matrices are given by
\begin{equation}
  \label{eq:35}
  \mat M_{\mathrm{PS}} =\begin{pmatrix}\ZB&0&0\\0&\ZS&0\\0&0&\ZM\end{pmatrix}\eqspace,
  \quad
  \mat V_{\mathrm{PS}}=\begin{pmatrix}
    0&    \e^{-\beta\VBS}&    \e^{-\beta\VBM}\\
    \e^{-\beta\VSB}& 0 &    \e^{-\beta\VSM}\\
    \e^{-\beta\VMB}&    \e^{-\beta\VMS}&    0
  \end{pmatrix}\eqspace,
  \quad
  \vec v =\begin{pmatrix}1\\1\\1\end{pmatrix}
  \eqspace.
\end{equation}
$\mat 1$ is the unity matrix. Evaluating \eq~\eqref{eq:34} yields the grand canonical
partition function of the three-state model
\begin{equation}
  \label{eq:36}
  \begin{split}
    \Z&=\biggl[ \ZB + \ZS + \ZM +\left(\e^{-\beta \VBS} +\e^{-\beta \VSB}\right) \ZB \ZS
    +\left(\e^{-\beta \VSM} +\e^{-\beta \VMS}\right) \ZS \ZM
    +\left(\e^{-\beta \VMB } +\e^{-\beta  \VBM  } \right)\ZM \ZB\\
    & \eqindent+\Bigl( \e^{-\beta ( \VBS + \VSM )} 
    +\e^{-\beta ( \VSM + \VMB )} 
    +\e^{-\beta ( \VMB + \VBS )} 
    +\e^{-\beta ( \VBM + \VMS )} 
    +\e^{-\beta ( \VMS + \VSB )} 
    +\e^{-\beta ( \VSB + \VBM )} 
    \\ & \eqindent\eqindent -\e^{-\beta ( \VBS + \VSB )} 
    -\e^{-\beta ( \VSM + \VMS )} 
    -\e^{-\beta ( \VMB + \VBM )} 
    \Bigr) \ZB \ZS \ZM
    \biggr]\\
    & \eqindent\times \biggl[ 1 -\e^{-\beta ( \VBS + \VSB )} \ZB \ZS -\e^{-\beta ( \VSM +
      \VMS )} \ZS \ZM
    -\e^{-\beta ( \VMB + \VBM )} \ZM \ZB\\
    & \eqindent\eqindent-\Bigl(\e^{-\beta ( \VBS + \VSM + \VMB )} 
    +\e^{-\beta ( \VBM + \VMS+ \VSB )}\Bigr) \ZB \ZS \ZM \biggr]^{-1} \eqspace,
  \end{split}
\end{equation}
which depends on the grand canonical partition functions~$\Z_i$ of the different regions.

\section{Origin of the logarithmic loop entropy}
\label{sec:orig-logar-loop}

In the simplest case a polymer can be described as an ideal random walk with step length
$b$ in $d=3$ dimensions.  When speaking of polymers $b$ is also called \definition{Kuhn
  length}.  Let us first consider a one-dimensional random walk. The probability to be
after $N$ steps at point $ x = n b$ is given by the binomial distribution
\begin{equation}
  \label{eq:1}
  P(x) = \frac{N!}{((N+n)/2)!((N-n)/2)!} \cdot p^{(N+n)/2} \cdot(1-p)^{(N-n)/2} 
  = \binom{N}{(N+n)/2} \cdot p^{(N+n)/2} \cdot(1-p)^{(N-n)/2} 
  \eqspace,
\end{equation}
where $p=1/2$ is the probability to move to the right and $(1-p)$ the probability to move
to the left. By virtue of the central limit theorem the binomial distribution can be
approximated by the normal distribution if $N$ is large
\begin{equation}
  \label{eq:2}
  P(x) \simeq \frac{1}{\sqrt{N b^2 2 \pi}} \e^{- \frac12 \frac{x^2}{Nb^2}}
  \eqspace.
\end{equation}
Now, we consider a random walk in three dimensions, where there are $N/3$ steps in each
spatial direction. Since the steps in each of the three directions are independent from
each other the probability distribution for being at point $\vec R = (x,y,z)$ after $N$
steps is
\begin{equation}
  \label{eq:3}
  P(\vec R)  = P(x) P(y) P(z) = \frac{1}{(N b^2 2 \pi/3)^{3/2}} \e^{- \frac32
    \frac{R^2}{Nb^2}}
  \eqspace.
\end{equation}
The probability of a closed random walk, \ie a loop, is the probability to return to the
origin after $N$ steps
\begin{equation}
  \label{eq:4}
  P(0) =  \frac{1}{(N b^2 2 \pi/3)^{3/2}}
  \eqspace.
\end{equation}
The entropy difference between an ideal polymer forming a loop and an unconstrained
polymer is therefore given by
\begin{equation}
  \label{eq:5}
  \Delta S_{\mathrm{loop}}(N) = \kB \ln P(0) \sim \kB \ln N^{-\loopexponent{}}
  \eqspace,
\end{equation}
where we introduced the \definition{loop
  exponent}~$\loopexponent{}=\loopexponent{}_{\mathrm{ideal}} = d/2 = 3/2$
\cite{Gennes1979}.

If one considers now self-avoiding polymers the loop exponent increases,
$\loopexponent{}_{\mathrm{SAW}} = d\nu \simeq 1.76$ with $\nu \simeq 0.588$ in $d=3$
dimensions~\cite{Gennes1979}. Hence the entropic penalty for closing a loop
increases. However, helices emerging from the loop increase $\loopexponent{}$ further. In
the asymptotic limit of long helical sections, renormalization group predicts
$\loopexponent{}_l = d\nu + \sigma_l-l\sigma_3$ for a loop with $l$ emerging
helices~\cite{Duplantier1986,Kafri2000} where $\sigma_l=\varepsilon l(2-l) /16
+\varepsilon^2 l (l-2)(8l-21)/512 +\mathcal O (\epsilon^3)$ in an $\epsilon=4-d$
expansion.  One obtains $\loopexponent{}_1=2.06$ for terminal loops present in hairpin
structures and $\loopexponent{}_2=2.14$ for internal loops.

The same analysis holds if one considers different topologies, namely peeling off one
strand from the other starting from a nick. In this case one can show that
$\openendexponent{} = 0$ for an ideal polymer and $\openendexponent{} = 0.092$ for a self
avoiding polymer~\cite{Duplantier1986,Kafri2000}.

Application of force to loops also alters the value of the loop exponent, which has been
shown by \citet{Hanke2008}.


\section{Effect of parameters on features of the force-extension curve}
\label{sec:effect-param-feat}

Different parameters of the model are linked to certain features of the force extension
curve. The chemical potentials $\gio$ determine the critical forces (\fig~\ref{fig:2}),
the segment lengths~$\segmentlength_i$ affect the maximal extensions (\fig~\ref{fig:3}) of
each state and the off-diagonal $\Vij$ determine the cooperativity of the transition
(\fig~\ref{fig:4}).

\begin{nofloat}
  \centering
  \includegraphics{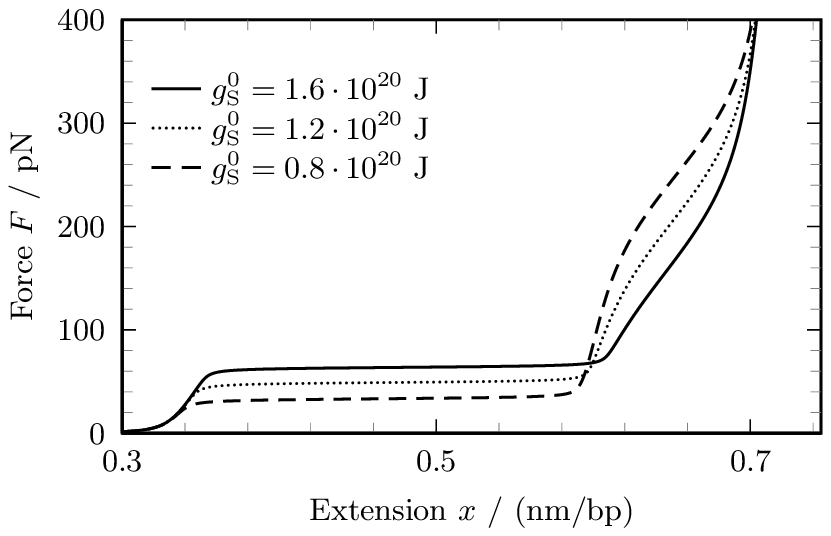}
  \captionof{figure}{The chemical potentials $\gio$ influence the transition forces. The
    solid line shows the force extension curve for the parameters used in the main text
    for DNA without DDP. The dashed and the dotted line show stretching curves with
    different chemical potentials of the \Sstate-state $\gSo$ while all other parameters
    are kept constant. Lowering $\gSo$ leads to a decreasing plateau force of the
    \Bstate\Sstate-transition as the free energy difference between \Bstate- and
    \Sstate-states ($\gS(\F) - \gB(\F)$) decreases, see \eq~\eqref{main-eq:2} in the main
    text. At the same time the force of the \Sstate\Mstate-transition increases as the
    free energy difference between \Sstate- and \Mstate-states ($\gM(\F) - \gS(\F)$)
    increases.  }
  \label{fig:2}
\end{nofloat}

\begin{nofloat}
  \centering
  \includegraphics{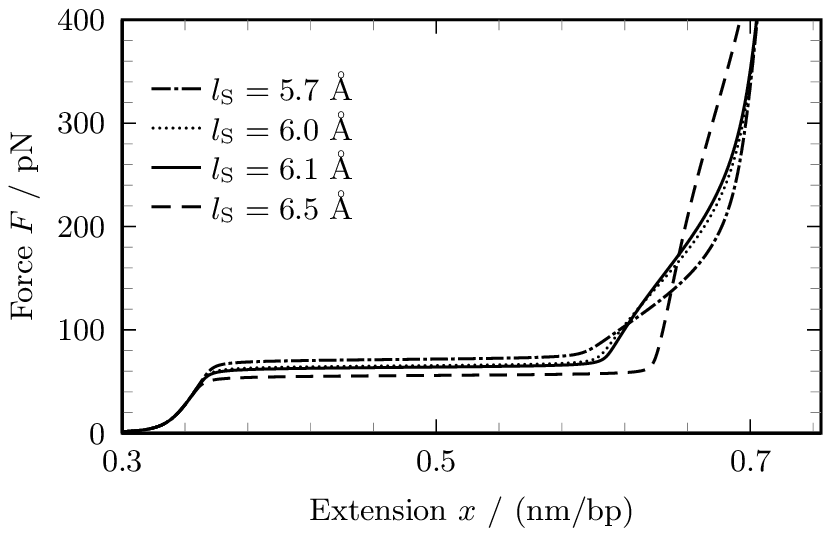}
  \captionof{figure}{The segment lengths $\segmentlength_i$ determine the extensions
    between which the transitions occur. The solid line shows the force extension curve
    for the parameters used in the main text for DNA without DDP. The other curves have
    been obtained by by varying the segment length of the \Sstate-state $\segmentlengthS$
    while keeping all other parameters constant. The effect of changing $\segmentlengthS$
    is twofold. First, increasing $\segmentlengthS$ elongates the
    \Bstate\Sstate-plateau. As a side effect a larger $\segmentlengthS$ stabilizes the
    \Sstate-state as its $\gSWLC(\F)$ is lowered.  This can also be concluded from
    \eq~\eqref{eq:33}. Thus, increasing $\segmentlengthS$ shifts the equilibrium towards
    the \Sstate-state, which is the same as decreasing $\gSo$, \cf \fig~\ref{fig:2}. }
  \label{fig:3}
\end{nofloat}

\begin{nofloat}
  \centering
  \includegraphics{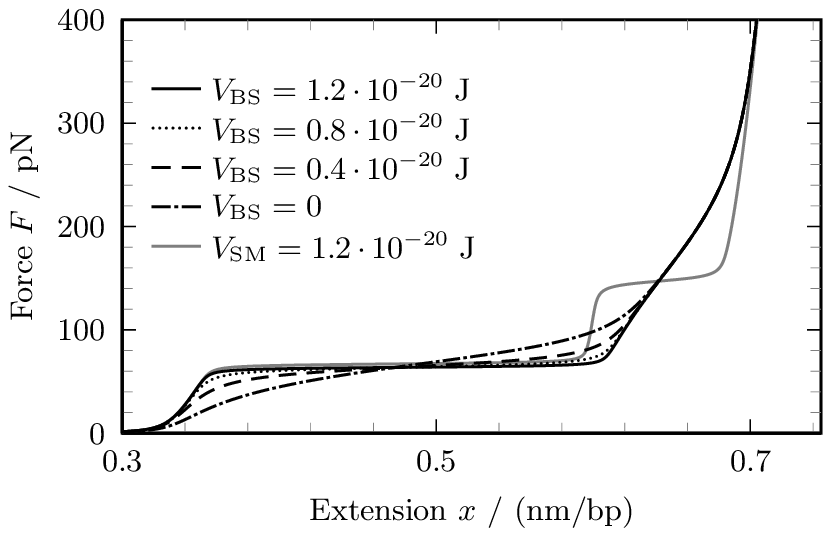}
  \captionof{figure}{The interfacial energies~$\Vij$ determine the cooperativity of the
    transitions. The solid line shows the force extension curve for the parameters used in
    the main text for DNA without DDP. The other force extension curves have been obtained
    by varying $\VBS$ or $\VSM$ while keeping all other parameters constant. Lowering
    $\VBS$ (dotted, dashed, and dash-dotted lines) smoothes the \Bstate\Sstate-transition
    and renders it less cooperative, which has been observed for DNA with DDP, \cf
    \fig~\ref{main-fig:2} in the main text and the experiments by
    \citet{Krautbauer2000}. $\VSM$ has similar effects on the melting transition:
    Increasing $\VSM$ from its original value $0$ to $\unit{1.2\cdot10^{-20}}{\joule}$
    (gray curve) sharpens the denaturation transition. }
  \label{fig:4}
\end{nofloat}

\begin{nofloat}
  \centering
  \includegraphics{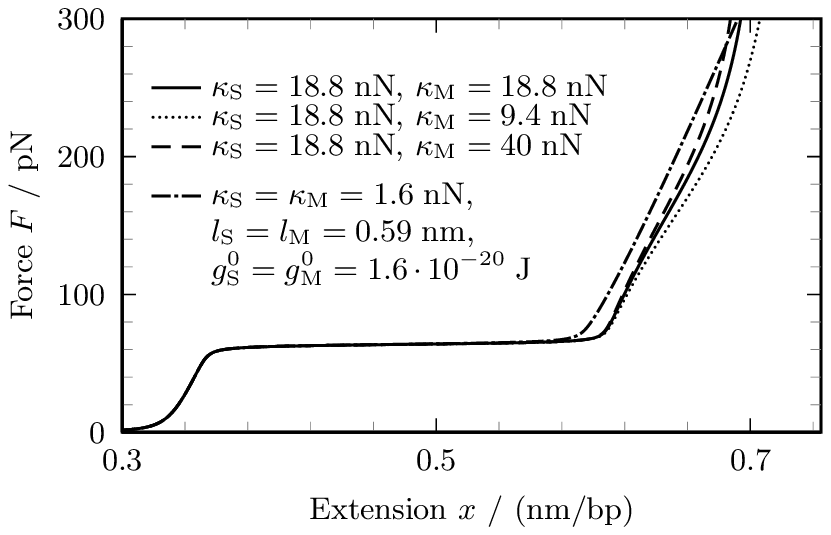}
  \captionof{figure}{The influence of the stretch modulus on force extension curves is
    quite moderate. The solid line shows the force extension curve for the parameters used
    in the main text for DNA without DDP.  The dashed and the dotted line show the data
    with $\stretchmodulusM = \unit{40}{\nano\newton}$ and $\stretchmodulusM =
    \unit{9.4}{\nano\newton}$, respectively, while keeping all other parameters
    unchanged. One observes that changing $\stretchmodulusM$ by a factor~$4$ results in
    deviations of the extension by just a few percent even at extreme forces
    $\F=\unit{400}{\pico\newton}$.  The dash-dotted line shows a force extension trace
    where $\stretchmodulusS = \stretchmodulusM = \unit{1.6}{\nano\newton}$, and
    $\segmentlengthS = \segmentlengthM = \unit{0.59}{\nano\meter}$ have been used, which
    are the typical experimental values for two parallel ssDNA \cite{Smith1996}. With this
    choice of parameters the force cannot discriminate between the \Sstate{}- and
    \Mstate{}-state anymore --~the two states are degenerate. Therefore we have to set the
    chemical potentials equal, fit them again and obtain $\gSo = \gMo =
    \unit{1.6\cdot10^{-20}}{\joule}$. Consequently the shoulder indicating a denaturation
    transition does not appear in the force extension curve in contrast to the other
    curves shown.  }
  \label{fig:1}
\end{nofloat}

\section{Tables with model parameters}\label{sec:tables-with-model}

\begin{nofloat}
  \centering
  \begin{tabular}{lcccl}
    Parameter & Symbol & $\lambda$-DNA & \text{$\lambda$-DNA + DDP} & Reference\\\hline
    segment length & $\segmentlengthB$ & 0.34 & 0.34 & \citet{Wenner2002}\\
    ($\nano\meter$)& $\segmentlengthS$ & 0.61 & 0.60 & fitted\\
    & $\segmentlengthM$ & 0.71 & 0.71 & \citet{Hugel2005}\\
    persistence length & $\persistencelengthB$ & 48 & 48 &
    \citet{Wenner2002}\\
    ($\nano\meter$)    & $\persistencelengthS$ & 25 & 25 & assumed\\
    & $\persistencelengthM$ & 3 & 3 & \citet{Murphy2004}\\
    stretch modulus & $\stretchmodulusB$ & 1.0 & 1.0 & \citet{Wenner2002}\\
    (\nano\newton) &$\stretchmodulusS $ & $2\cdot9.4$& $2\cdot9.4$ &\citet{Hugel2005}\\
    &$\stretchmodulusM $ & $2\cdot9.4$& $2\cdot9.4$ &\citet{Hugel2005}\\
    chemical potential & $\gBo$ & 0.0 & 0.0 & defined \\
    ($\unit{10^{-20}}{\joule}$)    & $\gSo$ & 1.6 & 1.2 & fitted\\
    & $\gMo$ & 2.4 & 2.8 & fitted\\
    interfacial energy & $\VBB, \VSS, \VMM$ & 0.0 & 0.0 & incorporated in
    $\gio$\\
    ($\unit{10^{-20}}{\joule}$)    & $\VBS$ & 1.2\textsuperscript{*} & 0.0 &     assumed, \textsuperscript{*}disrupt stacked base pairs \cite{SantaLucia1998}\\
    & $\VBM$ & 1.2\textsuperscript{*} & 0.0 &
    assumed, \textsuperscript{*}disrupt stacked base pairs \cite{SantaLucia1998}\\
    & $\VSM$ & 0.0 & 0.0 & 
  \end{tabular}
  \captionof{table}{Parameters for stretching curves at temperature
    $T=\unit{293}{\kelvin}$.}
  \label{tab:1}
\end{nofloat}

\begin{nofloat}
  \centering
  \begin{tabular}{lc@{\hspace{3ex}}cl}
    Parameter & Symbol & $\lambda$-DNA  & Reference\\\hline
    persistence length & $\persistencelengthB(T)$ & $48/(T/\unit{293}{\kelvin})$  &\\
    ($\nano\meter$) & $\persistencelengthS(T)$ & $25/(T/\unit{293}{\kelvin})$  &\\
    & $\persistencelengthM(T)$ & $3/(T/\unit{293}{\kelvin})$  & \\
    chemical potential & \multicolumn{2}{c}{$\gio(T) = h_i - T s_i$}&  \\
    enthalpy & $h_{\mathrm{\Bstate}}$ & $ 0.0 $ & \\
    ($\unit{10^{-20}}{\joule}$) & $h_{\mathrm{\Sstate}}$ & $ 7.0 $ & \citet{ClausenSchaumann2000},
    slightly rescaled \\
    & $h_{\mathrm{\Mstate}}$ & $ 15 $ & fit for $\loopexponent = 0$ \\
    & $h_{\mathrm{\Mstate}}$ & $ 16 $ & fit for $\loopexponent = 3/2$ \\
    entropy& $s_{\mathrm{\Bstate}}$ & $ 0.0 $ & \\
    ($\unit{10^{-22}}{\joule\per\kelvin}$) & $s_{\mathrm{\Sstate}}$ & $ 1.8 $ & \citet{ClausenSchaumann2000},
    slightly rescaled \\
    & $s_{\mathrm{\Mstate}}$ & $ 4.2 $ & fit for $\loopexponent = 0$ \\
    & $s_{\mathrm{\Mstate}}$ & $ 4.6 $ & fit for $\loopexponent = 3/2$ 
  \end{tabular}
  \captionof{table}{Temperature dependence of parameters for DNA without DDP. The
    enthalpy~$h_{\mathrm{\Mstate}}$ and entropy~$s_{\mathrm{\Mstate}}$ of the denatured
    state are determined by fixing the denaturation temperature~$\Tc$ for zero force at
    $\Tc = \unit{348}{\kelvin}$. }
  \label{tab:2}
\end{nofloat}


\bibliography{dna_BSL-bib}